\documentstyle[sprocl,epsf]{article}

\def\Pom{{\bf I\!P}}
\def\lsim{\mathrel{\rlap{\lower4pt\hbox{\hskip1pt$\sim$}}
    \raise1pt\hbox{$<$}}}         %less than or approx. symbol
\def\gsim{\mathrel{\rlap{\lower4pt\hbox{\hskip1pt$\sim$}}
    \raise1pt\hbox{$>$}}}         %greater than or approx. symbol

\def\beq{\begin{equation}}
\def\eeq{\end{equation}}
\def\bea{\begin{eqnarray}}
\def\eea{\end{eqnarray}}
\begin{document}
\rightline{{\bf\em FZ-IKP(TH)-1998-5}}
\vspace{1.0cm}
\title{ THE RUNNING BFKL: PRECOCIOUS ASYMPTOPIA FOR CHARM 
AND THE $dF_{2}/d\log Q^2$ - PUZZLE 
\footnote {Talk at 6th International Workshop on
 Deep Inelastic Scattering and QCD (DIS'98), Brussels, April 1998}}

%\author{A. B. AUTHOR, C. D. AUTHOR}

%\address{World Scientific Publishing Co, 1060 Main Street, 
%River Edge,\\ NJ 07661, USA\\E-mail: wspc@wspc.com} 

\author{V. R. ZOLLER}

\address{Institute for  Theoretical and Experimental Physics,\\
Moscow 117218, Russia\\E-mail: zoller@heron.itep.ru}

\maketitle\abstracts{The running BFKL equation gives rise to a series of
 moving poles in the complex $j$-plane.
 The  first nodes for all subleading solutions 
(color dipole cross sections) accumulate at 
$r_1\sim 0.1\,{\rm fm}$.
Therefore the processes dominated by the dipole sizes $r\sim r_1 $
 are free of subleading BFKL corrections. An example -
 the leptoproduction of charm. The calculated 
$F_2^{cc}$ is exhausted by the leading BFKL pole and
gives a perfect description of the experimental data.
The logarithmic slope of the subleading structure functions,
 $dF_2^{(n)}/d\log Q^2$, 
at small $Q^2$ is  small compared to  $dF_2^{(0)}/d\log Q^2$ due to the
presence of nodes. This observation provides an explanation for 
the observed $x$/$Q^2$ dependence
of the derivative of the proton structure function  $dF_2/d\log Q^2$.}

\section{Introduction}

The BFKL equation for the interaction cross section
$\sigma(\xi,r)$ of the color dipole $\vec r$ with the target reads
${\partial \sigma(\xi,r)/ \partial \xi} ={\cal K}\otimes
\sigma(\xi,r)$ (hereafter $\xi =\log(1/ x)$).
 The kernel ${\cal K}$ is related to the flux of the Weizs\"acker-Williams
 soft gluons $|\vec E(\vec\rho_1)-\vec E(\vec\rho_2)|^2$.
 The Asymptotic Freedom (AF)  dictates
that the  chromoelectric fields $\vec E(\vec\rho)$ be calculated with the running QCD
charge $g_S(r_m)=\sqrt{4\pi\alpha_S(r_m)}$ taken at shortest relevant distance
$r_m=min\{r,\rho\}$ and
 $\vec E(\vec\rho)=g_S(r_m){{\vec \rho}/\rho^2}\times ({\rm screening\, factor})$.
The lattice QCD suggests
the Yukawa screening radius 
$R_c\simeq 0.2-0.3\,{\rm fm}$ \cite{PISMA1,NZZJETP,PLNZ1}.  
The so  introduced  running coupling does not exhaust all 
NLO effects but correctly describes the crucial enhancement of long distance,
 and suppression of short distance, effects by AF.

Our principal findings on the runnig color dipole BFKL equation
 are as follows \cite{DIS97,JETP}.
The spectrum of the running BFKL equation is  a series of
 moving poles $\Pom_n$ in the complex $j$-plane with eigenfunctions
$\sigma_{n}(\xi,r)=\sigma_{n}(r)\exp(\Delta_{n}\xi)$ being a solution of
${\cal K}\otimes \sigma_{n}=\Delta_{n}\sigma_{n}(r)$.
  The leading eigenfunction $\sigma_0(r)$
 is node
free.  The  subleading  $\sigma_n(r)$
 has $n$ nodes.
 The intercepts
$\Delta_n$ 
 closely, to better than $10\%$, follow the law
$ \Delta_n= {\Delta_0/ (n+1)}$
suggested earlier by Lipatov \cite{LIPAT86} .
The intercept of the leading pole trajectory,
with the above specific choice of
 $R_c$, is $\Delta_0\equiv\Delta_{\Pom}=0.4$.
The subleading $\sigma_{n}$  
 are  similar \cite{DIS97,JETP}  to Lipatov's quasi-classical
 eigenfunctions \cite{LIPAT86}.
Within our specific choice of $R_c$
 ( infrared regularization ) the  node of $\sigma_{1}(r)$
is located at $r=r_1\simeq 0.05-0.06\,{\rm fm}$, 
for larger $n$ the first node
moves to a somewhat  larger $r\sim 0.1\, {\rm fm}$.
 Hence, 
 $\sigma(\xi,r_1)$ is dominated by $\sigma_0(\xi,r_1)$ and
exhibits the precocious  BFKL asymptotics \cite{PLNZ2}.
Consequently, zooming at   $\sigma(\xi,r_1)$ one can readily
 measure  $\Delta_{\Pom}$ and such a zooming is possible in charm production.
\begin{figure}
\epsfxsize=0.4\hsize
\epsfbox{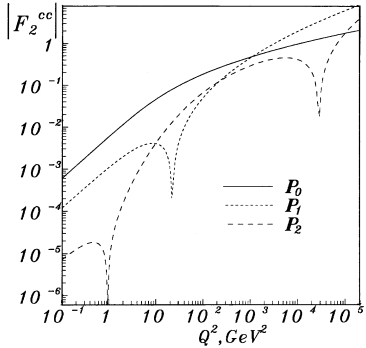}
\caption{Modulus of the  charm  structure functions $F_2^{(n)}$
for the BFKL poles $\Pom_n$ with n=0,1,2\, .}
\label{fig1}
\end{figure}
%..................................................
\section{ Nodal structure of BFKL solutions as
 seen by $c\bar c$-scanner}
\begin{figure}
\epsfxsize=0.7\hsize
\epsfbox{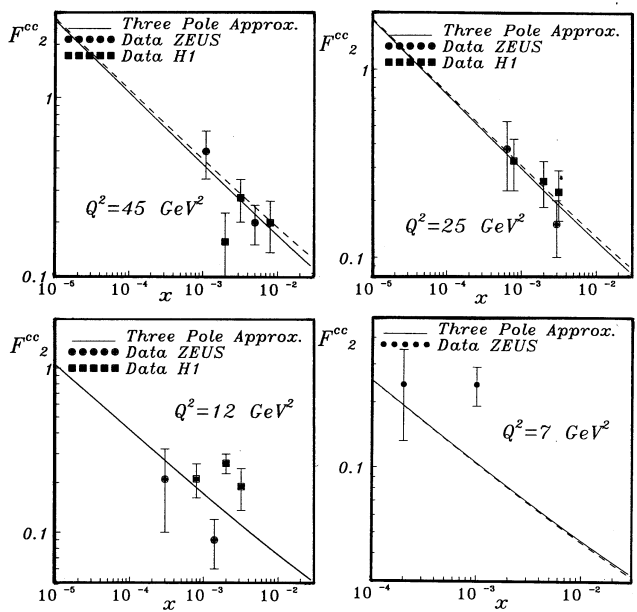}
\caption{The predicted  charm structure functions vs.
 H1   and ZEUS   data. The contribution of the leading pole with 
$\Delta_{\Pom}=0.4$ is shown by dashed line.}
\label{fig2}
\end{figure}

The color dipole representation for the  charm structure
 function  \cite{NZ91}
\beq
F_2^{cc}(x,Q^2)=
{Q^2\over {4\pi\alpha_{em}}}\int_0^1 dz\int 
d^2\vec r|\Psi_{cc}(z,r)|^2\sigma(x,r)\,,
\label{eq:F2CC}
\eeq
in conjunction with the explicit form
of the $c\bar c$ light-cone wave function, 
$\Psi_{cc}(z,r)$, shows that $F_2^{cc}(x,Q^2)$ is dominated by
 $r\sim r_{cc}\sim 1/\sqrt{4m_c^2+Q^2}$, and for a broad range of $Q^2$
one has $r_{cc}\sim r_1$.
 The BFKL-Regge
 expansion 
\beq
\sigma(x,r)=\sigma_0(r)(1/x)^{\Delta_0}+\sigma_1(r)(1/x)^{\Delta_1}+
\sigma_2(r)(1/x)^{\Delta_2}+...\,.
\label{eq:sigma}
\eeq
gives the BFKL-Regge expansion for the structure function
\beq
F_2^{cc}(x,Q^2)=\sum_nF_2^{(n)}(Q^2)(1/x)^{\Delta_n}\,.
\label{eq:F2BFKL}
\eeq
Because of the node of $\sigma_n(r)$
 the subleading charm structure functions  are  
negligible compared to $F_2^{(0)}$ in a broad range of $Q^2$ (Fig.\ref{fig1} ).
In Fig. \ref{fig2} our predictions for the charm structure 
function are compared with data from
H1 \cite{H1cc} and ZEUS \cite{ZEUScc}.
We correct for threshold effects  by the rescaling \cite{Barone}
$x\to x(1+4m_c^2/Q^2)$. 
From both Fig.\ref{fig1} and Fig.\ref{fig2} it is clear
that
the charm production for $Q^2\lsim 100\,{\rm GeV^2}$
 provides the unique opportunity of getting hold of elusive BFKL
asymptotics and measuring $\Delta_{\Pom}$
 already  at currently available $x$, $Q^2$.  

\section{ Running BFKL resolves 
  $dF_2/d\log Q^2$ - puzzle}
Caldwell's presentation \cite{DERLOG}
 of the HERA data in terms of $dF_2/d\log Q^2$
for
the all-flavor proton structure function $F_2(x,Q^2)$
 exhibits the  turn-over of the slope towards small $x$ and/or $Q^2$
in a striking contrast with the conventional DGLAP expectations \cite{GRV}.
Running BFKL resolves the puzzle. In  $F_2$
the contribution of the subleading poles is substantial. However,
because the all-flavor subleading $F_2^{(n)}(Q^2)$
 have node at $Q^2\sim Q^2_1\sim 10\,{\rm GeV^2}$ \cite{DIS97,JETP},
 their contribution
to the slope  $dF_2/d\log Q^2$ is negligible at $Q^2\lsim Q^2_1$.
Hence  $dF_2/d\log Q^2$ at small $Q^2$ follows closely $dF^{(0)}_2/d\log Q^2$.
Numerically,  at small $Q^2$,
 $F_2^{(0)}\sim Q^2/(\Lambda^2+Q^2)$ with $\Lambda^2\simeq 0.9\,{\rm GeV^2}$.
Therefore,
 $dF^{(0)}_2/d\log Q^2$ rises with $Q^2$ up to $Q^2\sim 1 {\rm GeV^2}$
then levels off.
Only at $Q^2\gsim Q^2_1$ when the subleading terms enter the game
$dF_2/d\log Q^2$ decreases and even becomes negative valued at
  large $x$.
\begin{figure}
\epsfxsize=0.5\hsize
\epsfbox{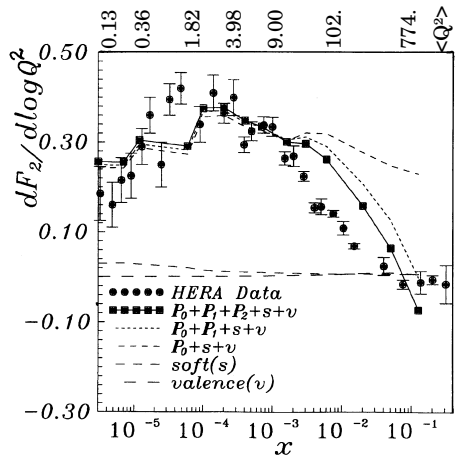}
\caption{Caldwell's plot of $dF_2/d\log Q^2$.
 Our  prediction is shown by 
the  solid line marked by squares.
 Notice the  non-monotonous dependence  of the slope on $x$
because the values of $<Q^2 >$ 
do not vary monotonously from one point to another.} 
\label{fig3}
\end{figure}
%............................................................

\end{document}